# Encoding Distortion Modeling For DWT-Based Wireless EEG Monitoring System

Alaa Awad, Medhat H. M. Elsayed, and Amr Mohamed
Department of Computer Science, Faculty of Engineering, Qatar University, Doha, Qatar
E-mail: aawad; mhamdy; amrm@qu.edu.qa

***Abstract*—Recent advances in wireless body area sensor networks leverage wireless and mobile communication technologies to facilitate development of innovative medical applications that can significantly enhance healthcare services and improve quality of life. Specifically, Electroencephalography (EEG)-based applications lie at the heart of these promising technologies. However, the design and operation of such applications is challenging. Power consumption requirements of the sensor nodes may turn some of these applications impractical. Hence, implementing efficient encoding schemes are essential to reduce power consumption in such applications. In this paper, we propose an analytical distortion model for the EEG-based encoding systems. Using this model, the encoder can effectively reconfigure its complexity by adjusting its control parameters to satisfy application constraints while maintaining reconstruction accuracy at the receiver side. The simulation results illustrate that the main parameters that affect the distortion are compression ratio and filter length of the considered DWT-based encoder. Furthermore, it is found that the wireless channel variations have a significant influence on the estimated distortion at the receiver side.**

***Index Terms*—Wireless healthcare applications, EEG signals, BASNs, Multiple Linear Regression, ANOVA.**

## I. INTRODUCTION

The rapid increase in the number of people that need continuous-remote healthcare monitoring, has increased the significance of electrocardiogram (ECG) and electroencephalogram (EEG) diagnosis systems. Advances in wireless communication systems and wearable sensors have facilitated implementing Body Area Sensor Networks (BASNs) as a promising technology that could meet this growing demand, and surpassing opportunity for ubiquitous remote real-time healthcare monitoring without constraining the activities of the patient [1]. Wireless BASNs consist of small sensor nodes in, on, or around a human body to monitor a range of vital signs which can be typically at a low data rate (e.g., body pressure, temperature or heart-rate reading), or at higher data rates such as streaming of EEG signals. These sensor nodes periodically transmit sensed data to a Health Monitoring Server.

The traditional wireless sensor network technologies are typically bulky, power hungry and based on MAC protocols such as Zigbee/IEEE 802.15.4 and Bluetooth, which are inefficient for such BASNs applications [2]. Moreover, most of these technologies ignore cross-layer design which optimizes the performance by jointly considering multiple protocol layers. In the last few years, most of the research work in the area of BASN has focused on issues related to sensors design, sensor miniaturization, signal compression techniques and efficient low-power hardware designs [3][4][5][6]. A good review of the state-of-the-art technologies, hardware, and standards for BASN was presented in [7].

In this paper, we focus on the electroencephalogram (EEG) applications. The EEG signal is considered as the main source of information to study human brain, which plays an important role in diagnosis of epileptic disease, brain death, tumors, stroke and several brain disorders [8]. EEG signals also play a fundamental role in Brain Computer Interface (BCI) applications [9]. To the best of our knowledge, the encoding distortion modeling of DWT-based compression as a function of compression ratio, wavelet filter length, and wireless channel variations has not been studied before. For example, the authors in [10] investigate the properties of compressed ECG data for reducing energy consumption using selective encryption mechanism and two-rate unequal error protection scheme. MAC (Media Access Control) layer energy-efficient design has also gained significant research interests recently by focusing on reducing power consumption at MAC layer by avoiding idle listening and collision [11], or by presenting latency-energy optimization [12]. The authors in [13] developed a MAC model for BASNs that fulfills the required reliability and latency of data transmissions, while maximizing battery lifetime of individual sensors. For that purpose, a cross-layer fuzzy-rule scheduling algorithm was introduced. However, the aforementioned work and the work in [14], they all ignored the encoding distortion modeling and did not take the encoding distortion nor the encoding energy into consideration in their model.

Accordingly, we propose an encoding distortion model of DWT-based compression that helps to anatomize, control, and optimize the behavior of the wireless EEG monitoring systems. Using this model, the encoder can dynamically update its complexity by reconfiguring its control parameters to maintain the energy constraints while retaining maximum reconstruction quality. Moreover, it can readily adapt to the channel dynamics, by updating the encoding parameters.

The rest of the paper is organized as follows. Section II describes the system model. Section III introduces the proposed encoder distortion model. Section IV discusses the effect of

This work was made possible by GSRA grant # GSRA2-1-0609-14026 from the Qatar National Research Fund (a member of Qatar Foundation). The findings achieved herein are solely the responsibility of the authors.

wireless channel variations on the estimated distortion, and presents the performance evaluation results. Finally, Section V concludes the paper.

## II. System Model

In this paper, wireless EEG monitoring system, as shown in Figure 1, is considered. In this model, the EEG data is gathered from the patient using EEG Headset [15], then it is sent to the PDA (i.e., smartphone) that compresses the gathered data and forwards it to a Health Monitoring Server (HMS).

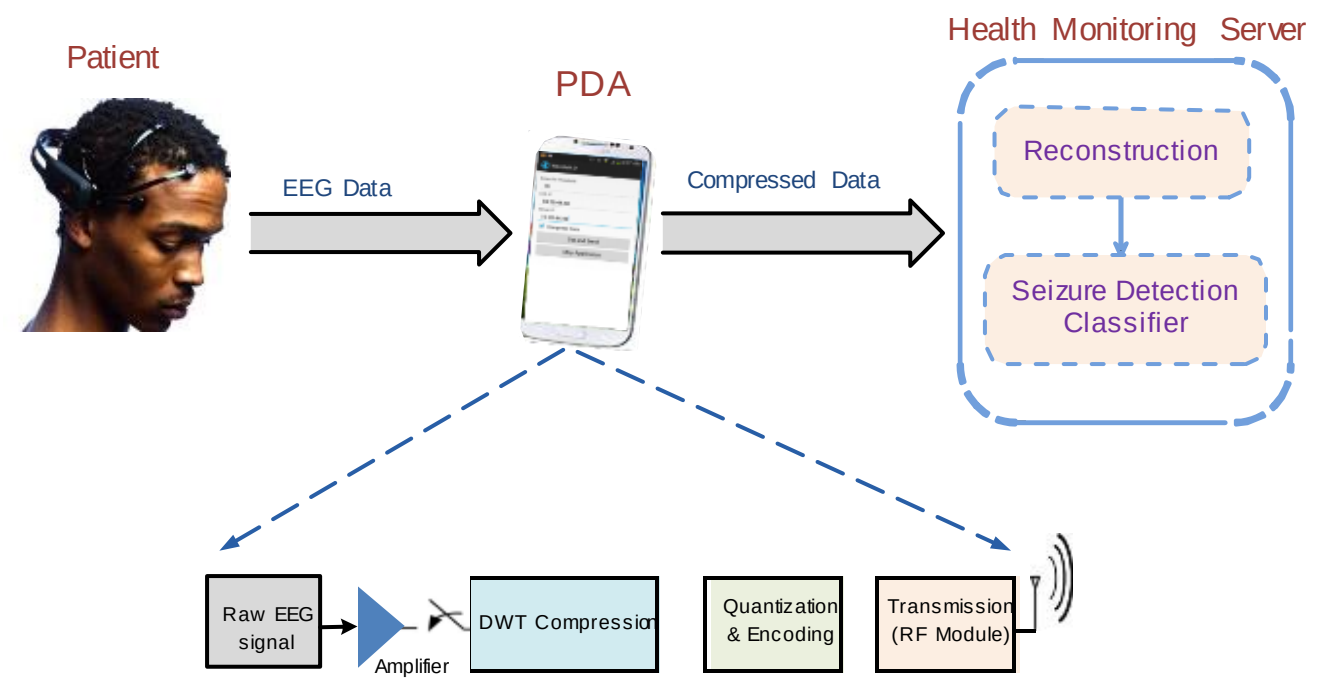

Fig. 1. System Model.

The general structure of the typical-used EEG encoder is illustrated in Figure 1. The main modules considered are amplifier and sampling, Discrete Wavelet Transform (DWT), quantization and encoding of the quantized DWT coefficients [16]. In what follows, we will derive and formulate the relation between the encoding distortion, the encoding parameters (i.e., wavelet filter length, compression ratio), and the transmission parameters (i.e., data length, transmission delay). First, it is assumed that the data will be transmitted over an ideal wireless channel. Afterward, we will exploit ANOVA analysis to study the effect of the imperfect channel on the estimated distortion, using four different channel models.

## III. Encoder Distortion Calculation

The encoding distortion is measured by the percentage Root-mean-square Difference (PRD) between the recovered EEG data and the original one, as follows

$$D_s = ((x - x_r) / x) * 100 \qquad (1)$$

where $x$ is the original signal and $x_r$ is the reconstructed signal. Using threshold-based DWT, the coefficients that are below the predefined threshold can be zeroed without much signal quality loss [17]. Accordingly, we control in the number of output samples generated from DWT, and thus the compression ratio of the DWT. The compression ratio is evaluated as

$$C_r = 1 - \frac{M}{N_s} \times 100 \qquad (2)$$

where $M$ is the number of output samples generated after DWT, and $N_s$ is the length of the original signal.

From our experimental results, the main parameters that affect the encoding distortion are the compression ratio ($C_r$), and the wavelet filter length ($F$) [17], as shown in Figure 2. Using source encoding, we can increase the compression ratio to reduce the amount of data to be transmitted, which in turn saves a significant amount of transmission energy. However, with increasing $C_r$, the transmitted data decreases, at the expense of increasing distortion. While with increasing $F$, more details is added to the sampled signal that leads to decrease the distortion (see Figure 3). These two conflicting objectives illustrate that in practical system design there is always a trade-off among the energy consumption, and the encoding distortion. Hence, it is crucial to develop an analytic framework to model the encoding distortion behavior of the EEG monitoring encoding system.

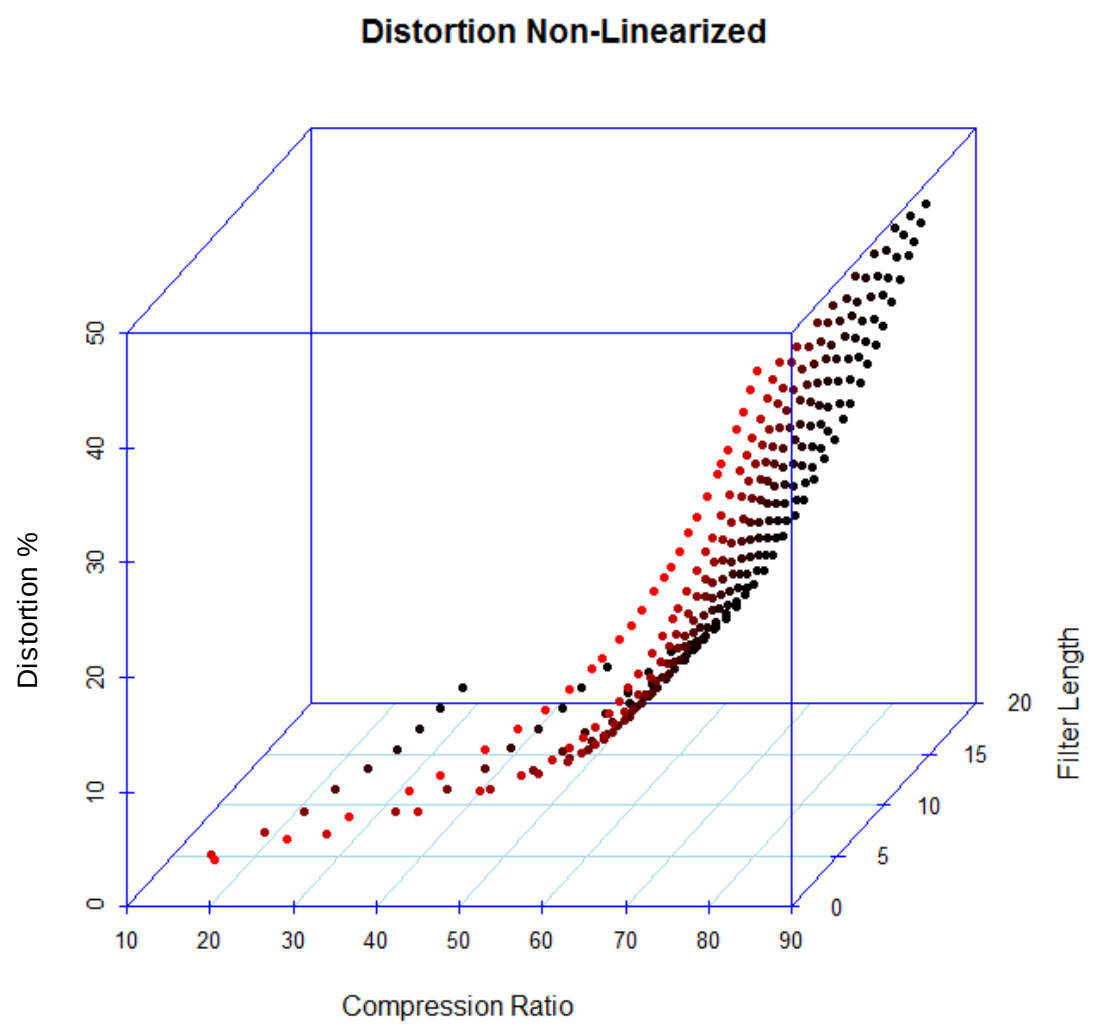

Fig. 2. The relation between the encoding distortion, the compression ratio and the wavelet filter length.

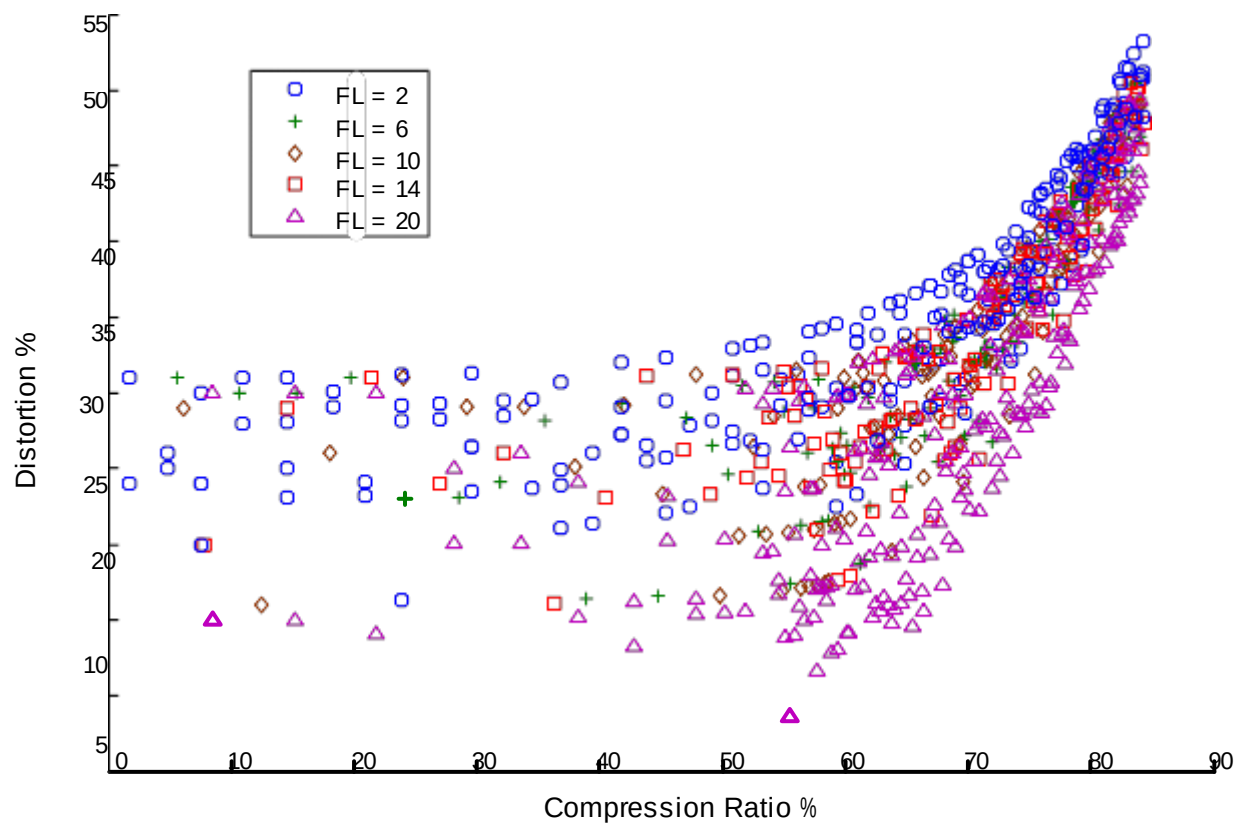

Fig. 3. Distortion variation with compression ratio for different filter length.

### A. Multiple Linear Regression

The regression model considers the relationship between encoding and transmission parameters on one hand and distortion on the other. However, this relation is non-linear, as shown in Figure 2. Thus, we use logarithmic transformation to convert the aforementioned relation into a linear relation, by considering $D = \log(D_s)$ (see Figure 4). As a result, the generated model will be as follow

$$D = \beta_0 + \beta_1 \cdot C_r + \beta_2 \cdot F + \beta_3 \cdot L + \beta_4 \cdot T \qquad (3)$$

where $L$ is the data length, and $T$ is the transmission delay.

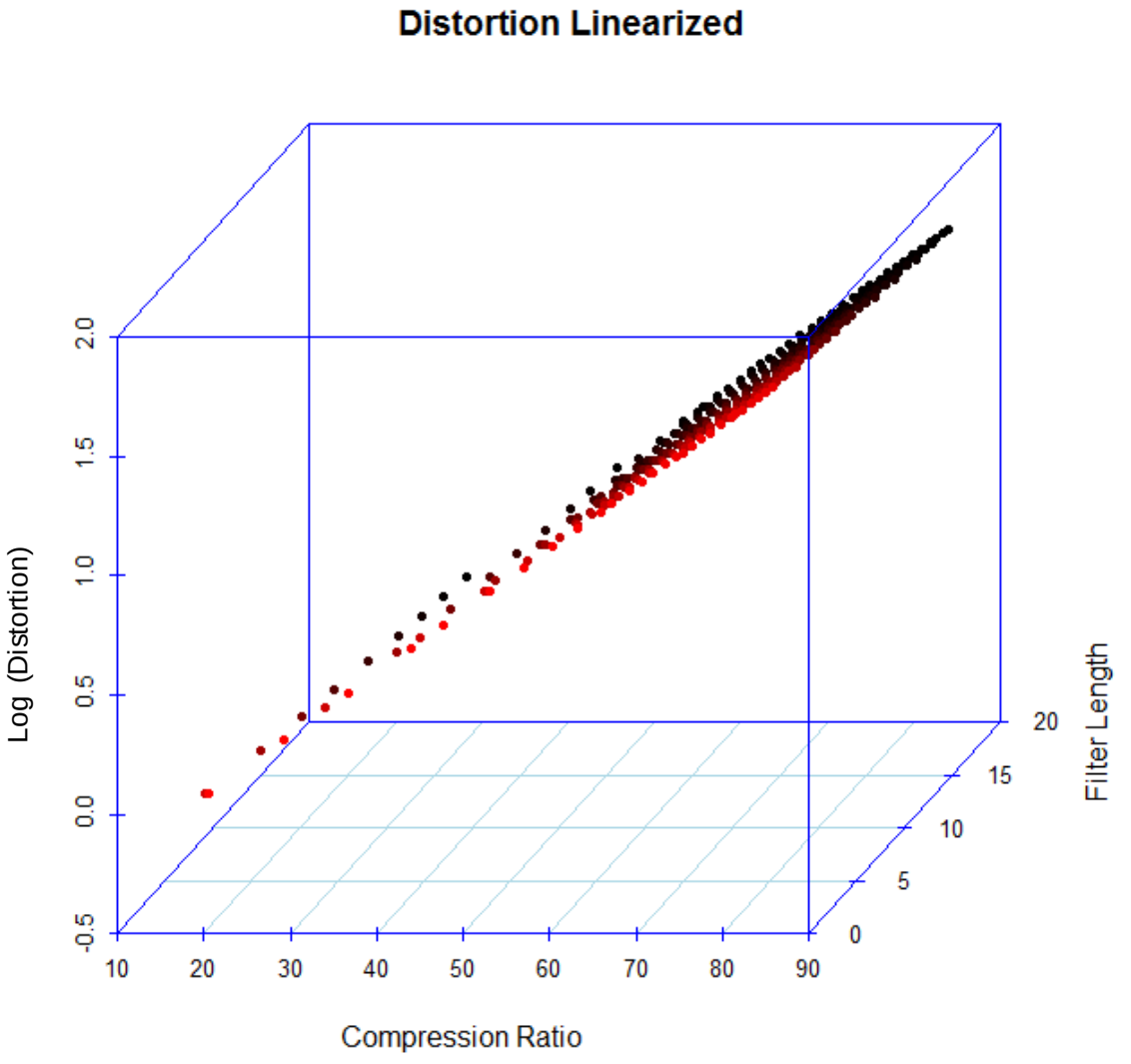

Fig. 4. Linearization of the encoding distortion.

We first present the the full model that considers all the parameters: wavelet filter length, compression ratio, data length, and transmission delay. Figure 5 provides a snapshot of the generated result for this model. The results show the coefficients and probabilities for each parameter. The coefcient of determination $R^2$, that indicates how well data points fit the line, has a high value of 0.955. It can be noticed also that the transmission delay parameter has high probability (i.e., $Pr(>|t|) = 0.938$), which means that this parameter has low contribution to the distortion.

### B. Model Selection

In order to reduce the complexity of the model, we begin to analyze the effect of reducing one of the parameters of this model. We conduct the regression analysis on a reduced model (i.e., reduced model 1) with ignoring the transmission delay parameter $T$ (see Figure 6). After that, to compare between the reduced model 1 and the full model, we apply ANOVA analysis [18]. The generated ANOVA result shows that the two models are similar with a probability 93.8%. Moreover, the $R^2$ parameter of the reduced model 1 has a high value of 0.9546.

```
> # Full Model
> # -----------
> FullModel = lm(LogDistortionIdeal~CompressionRatioIdeal+FilterLengthIdeal+DataLength+Transm$
> summary(FullModel)

Call:
lm(formula = LogDistortionIdeal ~ CompressionRatioIdeal + FilterLengthIdeal +
    DataLength + TransmissionDelay)

Residuals:
     Min       1Q   Median       3Q      Max
-0.13803 -0.04085 -0.00553  0.04479  0.39434

Coefficients:
                        Estimate Std. Error t value Pr(>|t|)
(Intercept)           -4.585e-01  2.874e-02 -15.955   <2e-16 ***
CompressionRatioIdeal  2.605e-02  3.314e-04  78.627   <2e-16 ***
FilterLengthIdeal     -8.142e-03  8.096e-04 -10.057   <2e-16 ***
DataLength            -5.018e-07  1.386e-06  -0.362    0.718
TransmissionDelay     -2.506e-04  3.234e-03  -0.077    0.938
---
Signif. codes:  0 '***' 0.001 '**' 0.01 '*' 0.05 '.' 0.1 ' ' 1

Residual standard error: 0.0787 on 293 degrees of freedom
Multiple R-squared:  0.9551,   Adjusted R-squared:  0.9545
F-statistic:  1557 on 4 and 293 DF,  p-value: < 2.2e-16
```

Fig. 5. Full regression model that includes filter length, compression ratio, data length, and transmission delay.

```
> # Reduced model 1 (remove TransmissionDelay)
> # ------------------------------------------
> ReducedModel1 = lm(LogDistortionIdeal~CompressionRatioIdeal+FilterLengthIdeal+DataLength)
> summary(ReducedModel1)

Call:
lm(formula = LogDistortionIdeal ~ CompressionRatioIdeal + FilterLengthIdeal +
    DataLength)

Residuals:
     Min       1Q   Median       3Q      Max
-0.13859 -0.04058 -0.00578  0.04426  0.39439

Coefficients:
                        Estimate Std. Error t value Pr(>|t|)
(Intercept)           -4.593e-01  2.686e-02 -17.100   <2e-16 ***
CompressionRatioIdeal  2.605e-02  3.308e-04  78.760   <2e-16 ***
FilterLengthIdeal     -8.132e-03  7.983e-04 -10.187   <2e-16 ***
DataLength            -5.052e-07  1.383e-06  -0.365    0.715
---
Signif. codes:  0 '***' 0.001 '**' 0.01 '*' 0.05 '.' 0.1 ' ' 1

Residual standard error: 0.07856 on 294 degrees of freedom
Multiple R-squared:  0.9551,    Adjusted R-squared:  0.9546
F-statistic:  2083 on 3 and 294 DF,  p-value: < 2.2e-16

>
> anova(FullModel, ReducedModel1)
Analysis of Variance Table

Model 1: LogDistortionIdeal ~ CompressionRatioIdeal + FilterLengthIdeal +
    DataLength + TransmissionDelay
Model 2: LogDistortionIdeal ~ CompressionRatioIdeal + FilterLengthIdeal +
    DataLength
  Res.Df    RSS Df   Sum of Sq     F Pr(>F)
1    293 1.8146
2    294 1.8146 -1 -3.7181e-05 0.006 0.9383
```

Fig. 6. Reduced model 1 that includes filter length, compression ratio, and data length.

It can be noticed also that the data length parameter can also be neglected (since $Pr(>|t|) = 0.715$). Therefore, we propose another reduced model (i.e., reduced model 2) that only considers the filter length and the compression ratio (see Figure 7). We can see that the similarity between this model and the original one is high, hence we can consider the reduced model 2 as an acceptable model. Accordingly, the

```
> # Reduced model 2 (remove DataLength)
> # ----------------------------------
> ReducedModel2 = lm(LogDistortionIdeal~CompressionRatioIdeal+FilterLengthIdeal)
> summary(ReducedModel2)

Call:
lm(formula = LogDistortionIdeal ~ CompressionRatioIdeal + FilterLengthIdeal)

Residuals:
     Min       1Q   Median       3Q      Max
-0.13798 -0.04047 -0.00559  0.04399  0.39253

Coefficients:
                        Estimate Std. Error t value Pr(>|t|)
(Intercept)           -0.4637498  0.0238932  -19.41   <2e-16 ***
CompressionRatioIdeal  0.0260596  0.0003300   78.96   <2e-16 ***
FilterLengthIdeal     -0.0081453  0.0007963  -10.23   <2e-16 ***
---
Signif. codes:  0 '***' 0.001 '**' 0.01 '*' 0.05 '.' 0.1 ' ' 1

Residual standard error: 0.07845 on 295 degrees of freedom
Multiple R-squared:  0.955,     Adjusted R-squared:  0.9547
F-statistic:  3134 on 2 and 295 DF,  p-value: < 2.2e-16

>
> anova(FullModel, ReducedModel2)
Analysis of Variance Table

Model 1: LogDistortionIdeal ~ CompressionRatioIdeal + FilterLengthIdeal +
    DataLength + TransmissionDelay
Model 2: LogDistortionIdeal ~ CompressionRatioIdeal + FilterLengthIdeal
  Res.Df    RSS Df   Sum of Sq      F Pr(>F)
1    293 1.8146
2    295 1.8154 -2 -0.00086032 0.0695 0.9329
```

Fig. 7. Reduced model 2 that includes filter length, and compression ratio.

relationship between the log of the distortion, filter length, and compression ratio can be modeled as a linear relation represented by equation 4.

$$D = \beta_0 + \beta_1 \cdot C_r + \beta_2 \cdot F. \quad (4)$$

This equation represents the relation between the log of the distortion and the compression ratio and filter length, where table I presents the equation parameters derived from the regression analysis.

TABLE I
REGRESSION PARAMETERS FOR ACCEPTABLE MODEL

| Regression Parameter | Value |
|---|---|
| $\beta_0$ | -0.46375 |
| $\beta_1$ | 0.02606 |
| $\beta_2$ | -0.0081453 |

## IV. PERFORMANCE EVALUATION

In this section, the effect of wireless channel variations on encoding distortion has been studied using R-Language simulation. The simulation results were generated using the network topology shown in Figure 1. Data acquisition is carried out using a MATLAB code that communicates with an Emotive device [19] to gather the data. We sensed 32 seconds that corresponds to 4096 samples of epileptic EEG data [20]. Then, the PDA gathers and forwards this data to the receiver side. At receiver side, the EEG feature extraction, classification and distortion evaluation are executed.

### *A. Channel Effect on Estimated Distortion*

In the previous sections, we assumed that the data will be transmitted from PDA to HMS through an ideal channel. This means that the channel will not add any distortion to the transmitted data. However, in the practical systems, the channel variations have an influence on the transmitted data. As a result, it is expected that the estimated distortion will increase due to channel impact. In this section, we consider four channel models: very good channel, good channel, bad channel, and very bad channel. As shown in Figure 8, the good channel model has a minor effect on the distortion, while the very bad channel has a major impact on the distortion. For the bad channel, most of the transmitted data is received incorrectly at the HMS, due to the channel effect, which leads to increase the estimated distortion at the receiver side (see Figure 8).

A more easy way to interpret the results is visualizing the channel effect on the distortion using box plot, as shown in Figure 9. Box plot is a graphical display showing center, spread, shape, and outliers. Accordingly, it can be seen that the different channel models have different influences on the distortion. As the channel goes worse, the distortion increase. To analysis the variance and study the differences among different channel models, we use ANOVA analysis, where the variable that defines group membership is *channel model*.

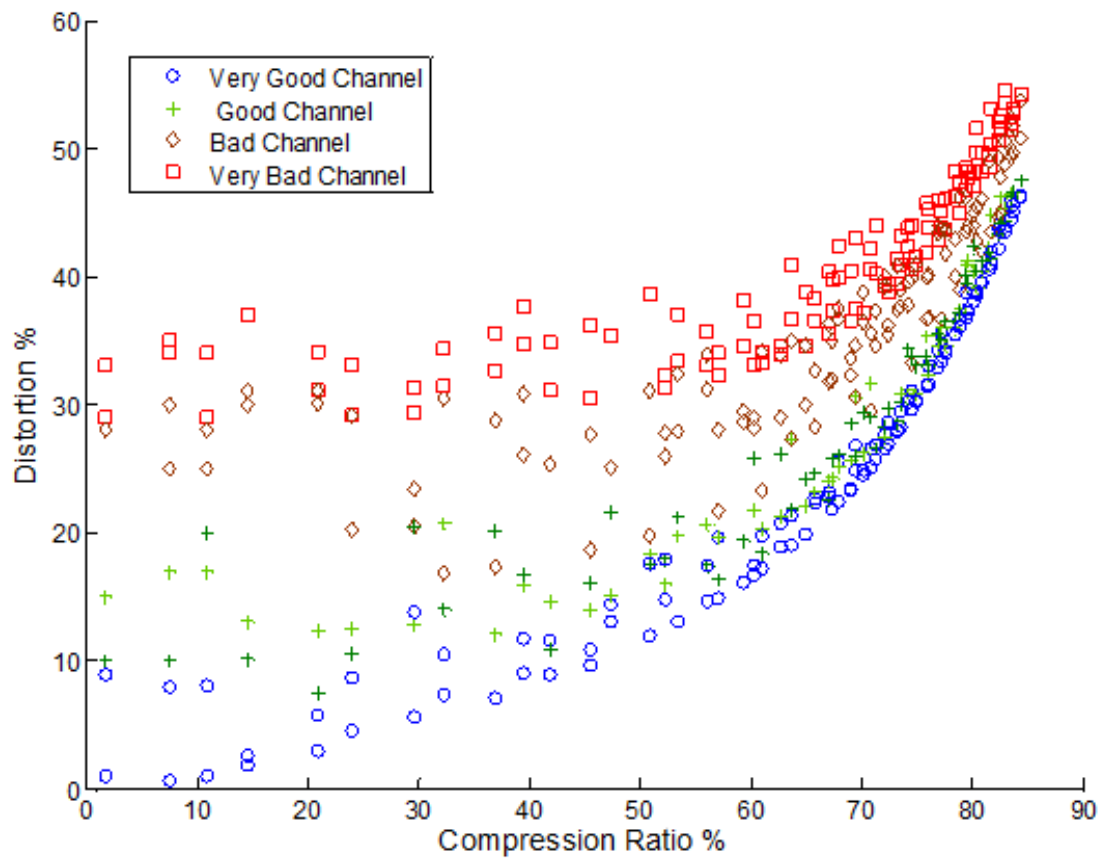

Fig. 8. Distortion variation with compression ratio for different channel models.

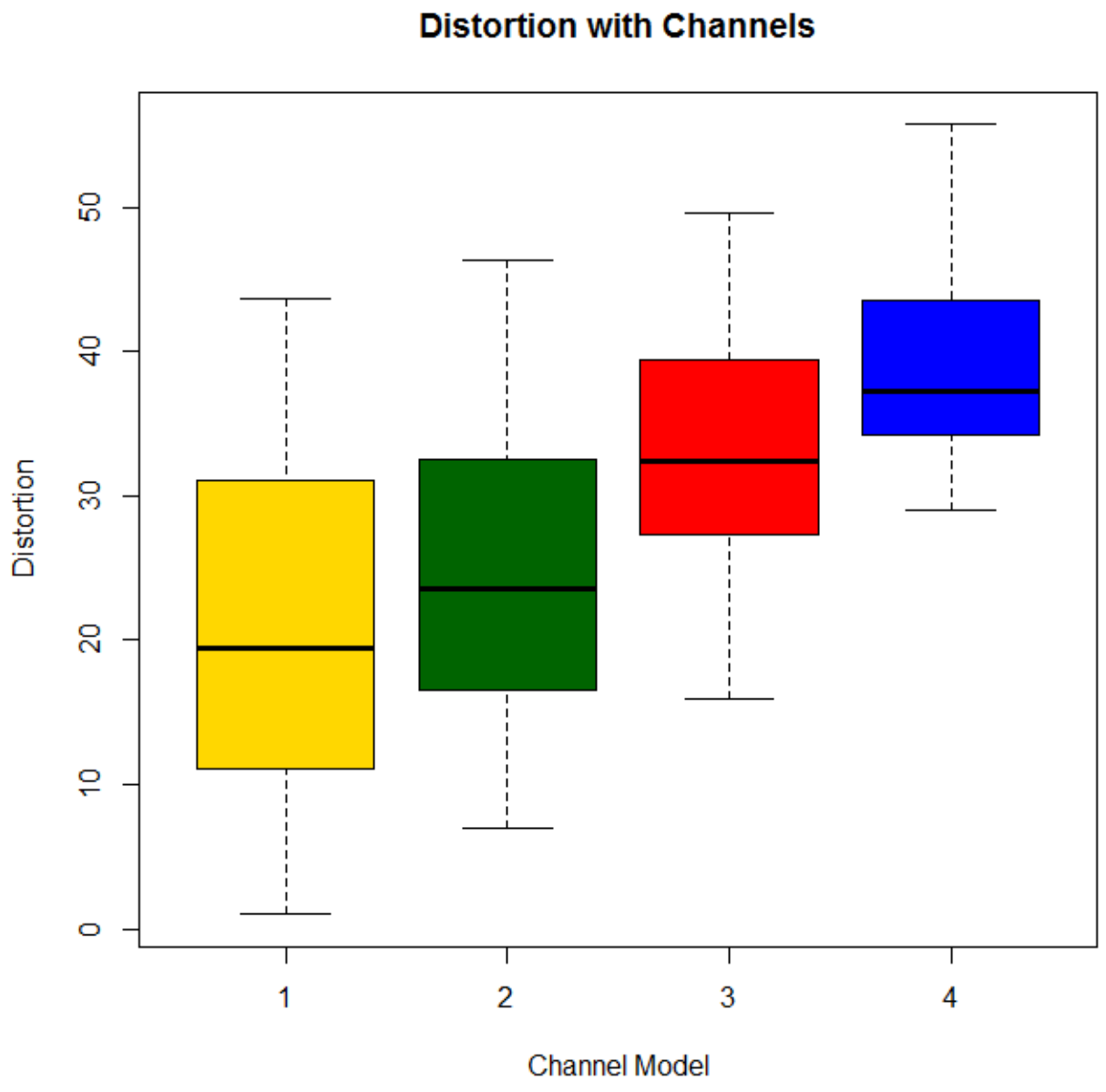

Fig. 9. Box plot of distortion variation with different channel models.

### *B. ANOVA analysis for real channel*

At the beginning, to know whether there is a difference between means of two samples of different channel models, we use hypothesis test on the mean difference [18], where we test the hypothesis that $H_0 : \mu_1 - \mu_2 = 0$, where $\mu_1$ and $\mu_2$ are the means of the samples of channel model 1 and 2, respectively. A snapshot of the used code is shown in Figure 10. In this context, we perform a hypothesis testing between channel model 1 and 2. From the results we can conclude that, type I error (i.e., Alpha error) has very low value, which indicates that the two channel models have a significant difference.

Since we have four channel models, analysis of variance is used to test for the differences among these models. This

```
> t Hypothesis testing for channels 1,
> t ---
> Meanl = ean(D1stort1on[Channel ==1))
> Mean2 = ean( Distort1On[Channel ==2 ))
>
>SOl = sd(Distort1on[Channel ==1))
> 502 = sd(Distortion[Channel ==2])
>
> nl = length(D1storuon [Channel ==1))
> n2 = length(Distortion[Channel =2 ))
>
> z = (Mean2- Meanl) /sqrt((SD1A2/nl) - (SD2A2/n2))
>
> Alpha = 2 • (1 - pnorm(z))     t 2-•1de hypothe•1s, Type I Error
>
> Alpha
[1]  4.H8104e-06
```

Fig. 10. Hypothesis testing between channel modell and 2.

analysis can be viewed as an extension of the t-test that we used for testing the means of two models. We conduct ANOVA analysis to compare the charmel models effects on the distortion, and to see the differences between these channel models. In order to fully compare the four charmel models conducted in our simulation, we perform ANOVA analysis as shown in Figure 11. The low probability acquired by the ANOVA analysis (i.e., *Pr(>* F)) shows that the four channel models are different. The stars also in Figure 11 indicate that there is a signicant chance for variance.

```
> t ANOVA Analy i
> t -    —    —    —
> anova (lm(LogDi storti.on-a3 .factor (Channel)))
Analysis of Variance Table

Response : LoqOi.s tort ion
                 Of  Sum Sq Mean Sq F value   Pr (>F)
as.!a.ctor( Channel)   3 21.019  7.0065  193.92  < 2.2e-16 ....
Residuals         1195 43.175  0.0361

Siqnif . codes: 0 ,,... , 0.001 ,..., 0.01 '*' 0.05 '.' 0.1 ' ' 1
> AnovaTable • aov (LogDi.st.or t.lon -a.s. faceor (Channe l))

> su.n:mary(AnovaTab e}
                    Of   Sum Sq Mean Sq F value   Pr (>F)
a: s.fact:or(Channe l)   3   21.02   7.006  193.9 <2e-16
Residuals          1195   43.18    0.036

Signif . codes : 0 ,,... , 0.001 ..... 0.01 ''' 0.05 '.' 0.1 ' ' 1
> TukeyHSD (AnovaTable )
   Tukey mul iple corr..parisons of rceans
     95\ tamily-wi se confidence level

Fit: aov (formula - LogDist.or ion - as.!actor (Channel ))

$ ·as.!actor (Channel )·
          ditt        lwr       upr   p adj
2-1 0.12377175 0.08381118 0.16373 23 o .oe+Oo
3-1 0.26660341 0.2 2661284 0.3065640  O.Oe+OO
4-1 0.31481381 0.30488327 0.38480 44  0.0e:+00
3-2 0.H283166  0.10290416 0.1827589  0.0e+00
1-2 0.22107209 0.18111189 0.2609993 0.Oe+OO
4-3 0.07824043 0.03831323 0.1181676  3.2e-06
```

Fig. 11. ANOVA analysis for the four channel models

The previous summary of ANOVA just indicates that there is a difference between the different charmel models, but does not say between which models. Hence, we used TukeyHSD() function to show the difference between each two models. In Figure 11, Column diff is giving the difference in the observed means, lwr is giving the lower end point of the interval, upr is giving the upper end point, and p-adj is giving the p-value after adjustment for the multiple comparisons. From p-adj column, we see that the probability of error if we reject hypothesis (i.e., p-value) is very low, that means we should reject the hypothesis that these models are similar. These results indicate that the differences between different models are significant. Moreover, an easy way to interpret the results of TukeyHSD is by visualizing the confidence intervals for the means differences, as shown in Figure 12.

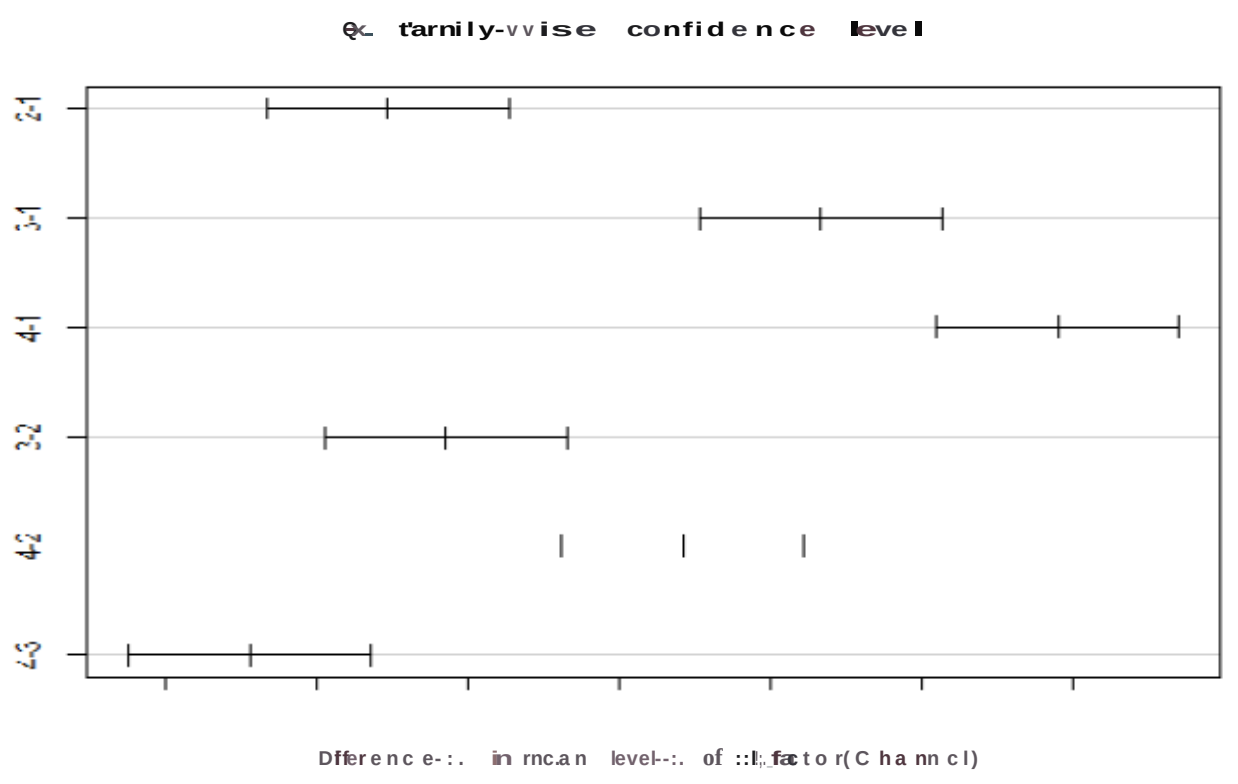

Fig. 12. TukeyHSD analysis for the four channel models

## V. CONCLUSION

In this paper, wireless EEG monitoring system is considered. In the proposed approach, using multiple linear regression, we modeled the encoding distortion of DWT-based compression as a function of compression ratio, wavelet filter length, data length, and transmission delay. Using this model, the encoder can effectively reconfigures its complexity by adjusting the control parameters to maintain the energy constraints while retaining maximum reconstruction quality. Furthermore, to reduce the complexity of the proposed model, we exploit model selection technique to decrease the number of irrelevant variables that have minor effects on the distortion. As a result, we found that the main parameters that affect distortion are compression ratio and wavelet filter length. In addition to that, the effect of wireless charmel variations on the estimated distortion has been studied. Using ANOVA analysis, we compared the effect of different channel models on the distortion. The simulation results has shown that the different charmel models have a significant influence on the estimated distortion at the receiver side.